\documentclass[]{article}
\usepackage[dvipdfmx]{graphicx}
\usepackage{
spconf,
amsmath,
amsfonts,
cite,
color,
mathrsfs,
multirow,
array,
fancyhdr,
}

\newcommand{\bm}[1]{\boldsymbol{#1}}

\newcommand{\mysection}[1]{
\vspace{-3pt}
\section{#1}
\vspace{-2pt}
}
\newcommand{\mysubsection}[1]{
\vspace{-7pt}
\subsection{#1}
\vspace{-2pt}
}

\newcommand{\mysubsubsection}[1]{
\vspace{-4pt}
\subsubsection{#1}
\vspace{-2pt}
}

\usepackage{algorithmic}
\usepackage{algorithm}

\usepackage{mathrsfs}

\newlength{\tablelength}
\newcolumntype{C}[1]{>{\centering\arraybackslash}p{#1\tablelength}}

\title{Data-driven design of perfect reconstruction filterbank \\ for DNN-based sound source enhancement
\vspace{-3pt}}
\name{Daiki Takeuchi$^\dagger$, Kohei Yatabe$^\dagger$, Yuma Koizumi$^\ddag$, Yasuhiro Oikawa$^\dagger$, Noboru Harada$^\ddag$
\vspace{-3pt}}
\address{
$^\dagger${\fontsize{11pt}{0pt}\selectfont Department of Intermedia Art and Science, Waseda University, Tokyo, Japan}\\
$^\ddag${\fontsize{11pt}{0pt}\selectfont NTT Media Intelligence Laboratories, Tokyo, Japan}
\vspace{-15pt}
}
\begin{document}
\ninept
\maketitle

\begin{abstract} 
\vspace{-3pt}
We propose a data-driven design method of perfect-reconstruction filterbank (PRFB) for sound-source enhancement (SSE) based on deep neural network (DNN).
DNNs have been used to estimate a time-frequency (T-F) mask in the short-time Fourier transform (STFT) domain.
Their training is more stable when a simple cost function as mean-squared error (MSE) is utilized comparing to some advanced cost such as objective sound quality assessments.
However, such a simple cost function inherits strong assumptions on the statistics of the target and/or noise which is often not satisfied, and the mismatch of assumption results in degraded performance.
In this paper, we propose to design the frequency scale of PRFB from training data so that the assumption on MSE is satisfied.
For designing the frequency scale, the warped filterbank frame (WFBF) is considered as PRFB.
The frequency characteristic of learned WFBF was in between STFT and the wavelet transform, and its effectiveness was confirmed by comparison with a standard STFT-based DNN whose input feature is compressed into the mel scale.
\vspace{-3.5pt}
\end{abstract}
\mysection{Introduction}
\label{sec:intro}

Sound-source enhancement (SSE) is used to recover the target sound from a noisy observed signal. 
A recent advancement of SSE is the use of a deep neural network (DNN) to estimate a time-frequency (T-F) mask in the short-time Fourier transform (STFT) domain 
\cite{Erdogan2015,weninger2014discriminatively,hershey2016icassp,Kolbaek2017,liu2017perceptually,Wang2018,Masuyama2019icassp}. 
In early studies, the mean-squared error (MSE) is used as the cost function to train the parameters of DNN \cite{xu2014, Erdogan2015,xu2015} because a gradient of MSE with respect to the parameters can be calculated analytically.
Among MSE-based costs, MSE between the target and masked signals on the complex plane, which is a cost function recently proposed for estimating a phase-sensitive mask (PSM) \cite{Erdogan2015}, is used as the comparison baseline system in many studies \cite{Kolbaek2017,williamson2017,Koizumi2018icassp}.

While simple cost functions such as MSE enables us to stably train DNN, they often inherit strong assumptions on the statistics of the target and/or noise.
For example, MSE assumes that the error of all frequency bins has zero means and uniform variance, which cannot be met in usual situations, unfortunately.
To overcome this problem, advanced cost functions, which directly increases performance measure of SSE, have been investigated such as the use of signal-to-distortion ratio (SDR) \cite{Venkataramani2017ACSSC} , Itakura-Saito divergence \cite{nugraha2016multichannel}, and objective sound quality assessments \cite{Koizumi2017icassp,fu2018,Koizumi2018assess}. 
These cost functions enable to increase the target performance directly; meanwhile, the complexity of gradient calculation is also increased and can result in unstable training.

At the same time, use of trainable T-F transforms has also been investigated \cite{sainath2015learning,Venkataramani2017ACSSC,Luo2018,wichern2018phase}.
These approaches utilized DNNs for transforming a signal into some domain similar to the T-F domain.
Their parameters were trained together with those of DNN for T-F mask estimation to improve the performance of SSE.
By simultaneously training the transformation, it can be possible to alleviate the assumption mismatch mentioned in the previous paragraph.
However, the trained transformation may not have the perfect reconstruction property, and increasing the number of DNN parameters increases the potential risk of over-fitting.
In addition, the transformed signals are less understandable, which might restrict the application.

An essence revealed by these studies on trainable T-F transformation is that the performance of SSE can be improved by training the T-F transform from a dataset.
That is, there is a room for improving SSE \textit{by designing a better T-F transform}.
The standard T-F transforms, including STFT and the wavelet transform, are well-understood as \textit{filterbank}, and its design strategy has been studied widely \cite{gan2003oversampled,cosentino2014cochlear,holighaus2014class,holighaus2015designing,muramatsu2017multidimensional,dam2017optimal}.
In particular, perfect-reconstruction filterbank (PRFB) is the important ingredient of SSE because the enhanced result must be converted back into the time domain%
\footnote{\textit{Perfect reconstruction} means that a signal which is not processed in the transformed domain can be perfectly reconstructed to the same signal in the original domain, i.e., no information loss happens by the transformation.}.
By utilizing one of those PRFBs, it should be possible to learn a T-F transformation which is understandable from both practical and theoretical point of views and performs better in terms of SSE.

\begin{figure}[t]
\centering 
\includegraphics[width = 0.95\columnwidth]{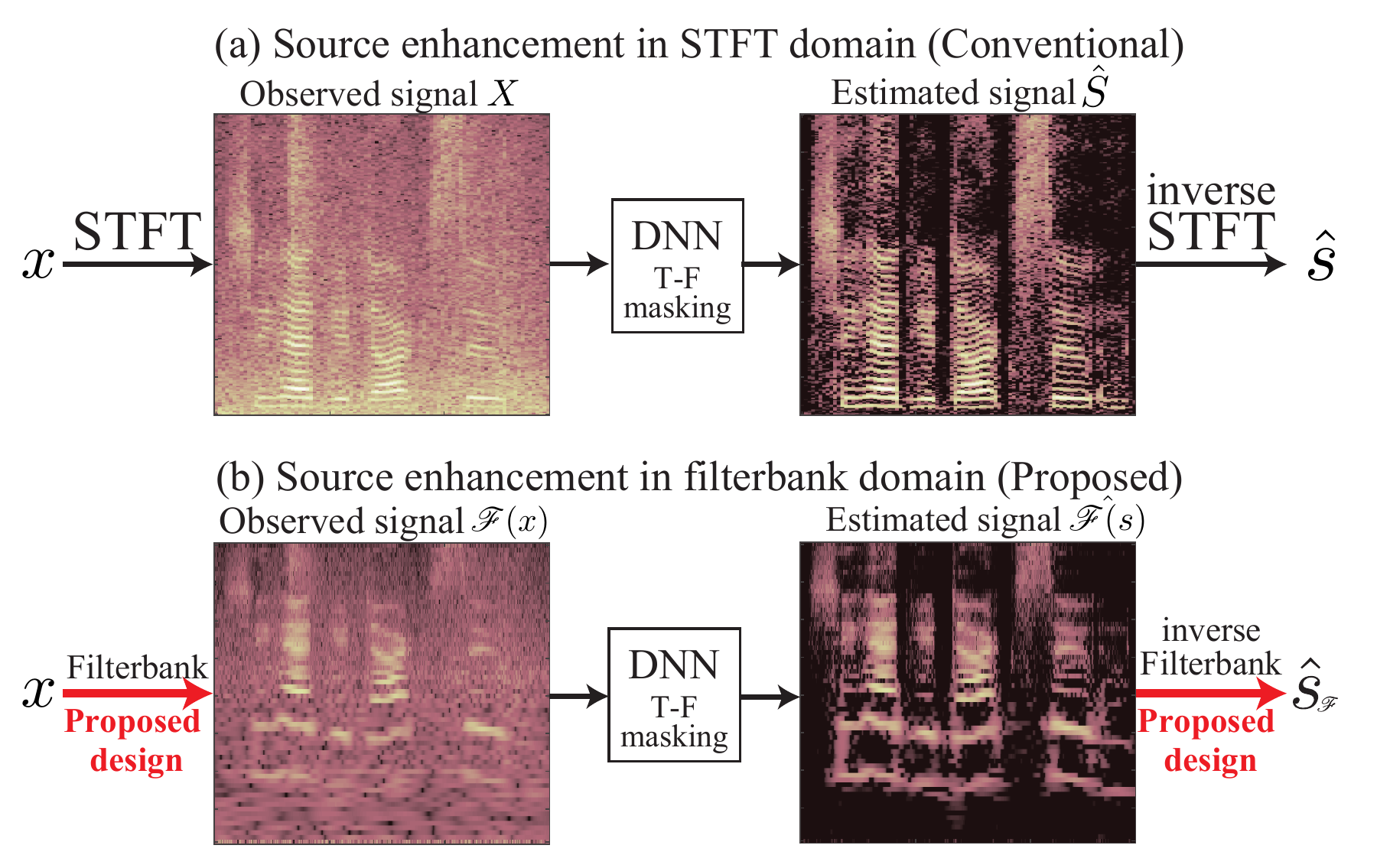}
\vspace{-10pt}
\caption{Illustration of the conventional and proposed method.}
\label{fig:propIm}
\vspace{-6pt}
\end{figure}

In this paper, we propose a method of designing PRFB for improving DNN-based SSE.
Our strategy is to design a filterbank from a dataset before training the DNN for T-F-mask estimation so that the training becomes easier.
The key idea is to compensate the mismatch of the assumption made by the cost function through the designed PRFB.
To do so, the warped filterbank frame (WFBF) \cite{holighaus2014class,holighaus2015designing} is utilized as PRFB, and its frequency-warping function is adapted to the database by calculating the amount of the error between clean and masked signals at each frequency.
As in Fig.~\ref{fig:propIm}, the designed WFBF (at the bottom) distributes the energy of a noisy signal more uniformly along the frequency than STFT (at the top), which should be an easier representation for MSE to be minimized.


\mysection{Time-frequency mask estimation}

The problem of SSE is to recover a target signal $s[t]$ degraded by noise $n[t]$.
An observed monaural signal $x[t]$ is modeled as
\begin{equation}
    x[t] = s[t] + n[t],
    \label{eq:sse_td}
\end{equation}
where $t$ is the time index.
STFT of a signal $\bm{x} = [x[1],\ldots,x[T]]^T$ with a window function~$\bm{g} \in \mathbb{R}^L$ is defined as
\vspace{-6pt}
\begin{equation}
\vspace{-6pt}
X[\omega,k]
= \sum\limits_{l = 0}^{L-1}
x[l-ak]\,
\overline{g[l] \, e^{2 \pi j \omega b l}},
\end{equation}
where $\overline{z}$ is complex conjugate of $z$, $j = \sqrt{-1}$, $a$ and $b$ are time and frequency shifting steps, and $\omega = 1,\ldots,\Omega$ and $k = 1,\ldots,K$ denote the frequency and time-frame indices, respectively.  
From the linearity, Eq.~\eqref{eq:sse_td} also becomes the summation in STFT domain 
\begin{equation}
    X[\omega,k] = S[\omega,k] + N[\omega,k].
    \label{eq:sse_STFTd}
\end{equation}
T-F masking is a standard method for SSE, where the estimated target signal $\hat{S}[\omega,k]$ is acquired by the element-wise multiplication of a T-F mask $G[\omega,k]$ to the observation $X[\omega,k]$ in the STFT domain: 
\begin{equation}
%
    \hat{S}[\omega,k] = G[\omega,k] \, X[\omega,k].
\end{equation}
Then, the output signal is transformed back to the time domain by the inverse STFT.
The T-F mask $G[\omega,k]$ must be estimated solely from $X[\omega,k]$, which is the difficult part of T-F masking.

\mysubsection{Deep learning for estimating phase sensitive mask (PSM)}
\label{subsec:DLmethod}

Many methods have applied deep learning for estimating the T-F mask.
In deep learning approach, a T-F mask $G[\omega,k]$ is estimated as 
\begin{equation}
\vspace{-1pt}
     \hat{G}[\omega,k]=  \mathcal{M}_{\theta}(\Psi)[\omega,k]
\end{equation}
where $\mathcal{M}_\theta$ is a regression function implemented by DNN, $\theta$ is a set of its parameters, and $\Psi$ is the input acoustic feature.

Typically, a T-F mask $G$ is chosen to be real-valued.
The truncated PSM $G_{\rm PSM}$ is one of the real-valued T-F masks which minimizes MSE between $\hat{S}[\omega,k]$ and ${S}[\omega,k]$ on the complex plane \cite{Erdogan2015}:
\begin{equation}
G_{\rm PSM}[\omega,k] =  
\mathcal{T}_{[0,1]}\! \left[
\frac{|S[\omega,k]|}{|X[\omega,k]|} 
\cos(\phi_{S[\omega,k]} - \phi_{X[\omega,k]})
\right],
\end{equation}
where $\mathcal{T}_{[a,b]} [ z ] = \min( \max(z, a), b )$ is the truncation operator, and $\phi_{S[\omega,k]}$ and $\phi_{X[\omega,k]}$ are phase angles of $S[\omega,k]$ and $X[\omega,k]$, respectively.
For approximating this mask by DNN $\mathcal{M}_{\theta}$, its parameters $\theta$ are trained to minimize the following MSE for all data in a dataset:
\vspace{-5pt}
\begin{equation}
\vspace{-5pt}
     \mathcal{J}_{\rm PSM}(\theta) = 
     \sum\limits_{\omega = 1}^{\Omega}
     \sum\limits_{k = 1}^{K}
     \left|
     \mathcal{M}_{\theta}(\Psi)[\omega,k] \, X[\omega,k] - {S}[\omega,k]
     \right|^2.
     \label{eq:usualMSE}
\end{equation}
The sigmoid function is often chosen as the activation function of the output layer of $\mathcal{M}_{\theta}$ in order to limit its value within $0$ to $1$.

\mysubsection{Weighted MSE for reducing assumption mismatch of MSE}

\begin{figure}[t]
\centering 
\includegraphics[width = 0.97\columnwidth]{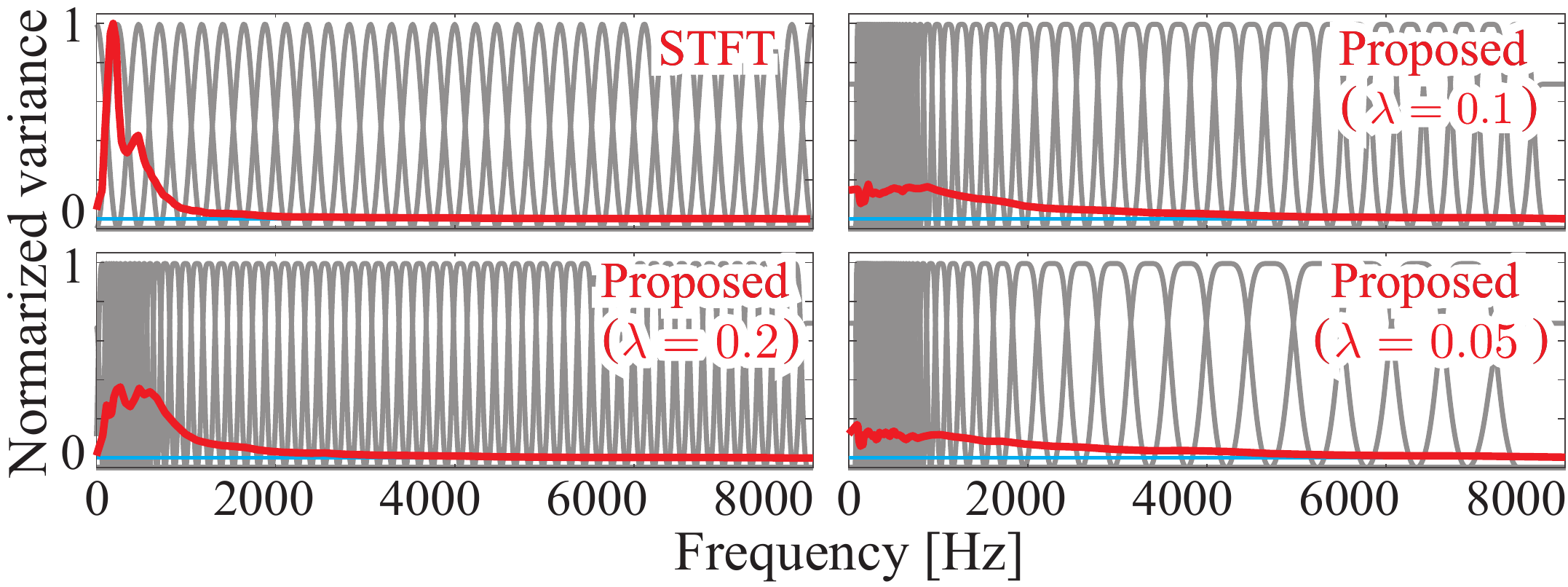}
\vspace{-5pt}
\caption{The variance of the oracle PSM masking error $\varepsilon$ (red) 
calculated in the STFT domain and the learned WFBF domains.}
\label{fig:preEx}
\vspace{-5pt}
\end{figure}

The above cost function, MSE, assumes that the error between the clean and masked T-F bin has the uniform variance for all bins.
However, this assumption cannot be met in reality because both target source and noise have non-uniform spectral distribution in practical situations.
Such assumption mismatch is problematic since it underestimates the error in the frequency range having small power.
That is, higher frequency range, which contains less power for practical sounds (see Fig.~\ref{fig:preEx}), is difficult to train than the lower range.

To normalize the error to make it uniform variance as the assumption of MSE, the cost function should be modified by weighting.
Since normalizing the time fluctuation is difficult as it is highly dependent on each signal, we consider a frequency-wise weighting,
\vspace{-2pt}
\begin{equation}
\vspace{-2pt}
     \mathcal{J}_{\rm WPSM}(\theta) = 
     \sum\limits_{\omega = 1}^{\Omega}
     \sum\limits_{k = 1}^{K}
     \left|
     W[\omega]
     (
     \mathcal{M}_{\theta}(\Psi)[\omega,k] \, X[\omega,k] - {S}[\omega,k]
     )     
     \right|^2,
     \label{eq:weightedLS}
\end{equation}
which results in the weighted MSE, where the weight is defined as the reciprocal of the frequency-wise standard deviation of the error,
\vspace{-1pt}
\begin{equation}
    W[\omega] = \Biggl \{
    \frac{1}{K}\sum\limits_{k = 1}^{K}\bigl(\varepsilon[\omega,k]\,\bigr)^2
    -
    \Biggl(\frac{1}{K}\sum\limits_{k = 1}^{K}\varepsilon[\omega,k]\Biggr)^{\!\!2}
    \,
    \Biggr \}^{\!\!-\frac{1}{2}},
\end{equation}
and the error is defined through the oracle PSM as
%
\begin{equation}
%
    \varepsilon[\omega,k] = 
    G_{\rm PSM}[\omega,k] \, X[\omega,k] - S[\omega,k].
    \label{eq:errorOraPSM}
\end{equation}
This weighted MSE equally treats the error for each frequency, which relieves the problem of the assumption mismatch of MSE because the weighted error has uniform variance for all frequencies.

However, the difficulty of training $\mathcal{M}_{\theta}$ cannot be completely removed by this weighting.
Although the cost function becomes more reasonable by the weighting, the optimization algorithm may not work appropriately for this function.
Since the difference between the maximum and minimum value of the weight $W[\omega]$ is typically large as the power for each frequency is highly unbalanced (see Fig.~\ref{fig:preEx}), the gradient becomes large for the frequency range having small power (since $W$ is reciprocal of power).
Then, the direction of the gradient becomes more sensitive to noise in the higher frequency range, and the effectiveness of a learning algorithm is reduced.
That is, the weighted MSE has a trade-off between the degree of assumption mismatch and difficulty of optimization.
From the optimization point of view, the weight $W[\omega]$ should be $1$ for all frequencies.


\mysection{Proposed method}
As discussed in the previous section, weighting is necessary for reducing the assumption mismatch, while the weight should not be utilized for stable optimization.
To resolve this dilemma, we propose to modify the T-F transform instead of modifying the cost function.

\mysubsection{Warped filterbank frame (WFBF) as PRFB}

To normalize frequency-wise error in MSE, we propose to use a frequency-warped PRFB so that the error for each frequency band has the same power.
For warping the frequency axis as desired, the WFBF \cite{holighaus2014class,holighaus2015designing} is considered in this paper.

\begin{figure}[t]
\centering 
\includegraphics[width = 0.97\columnwidth]{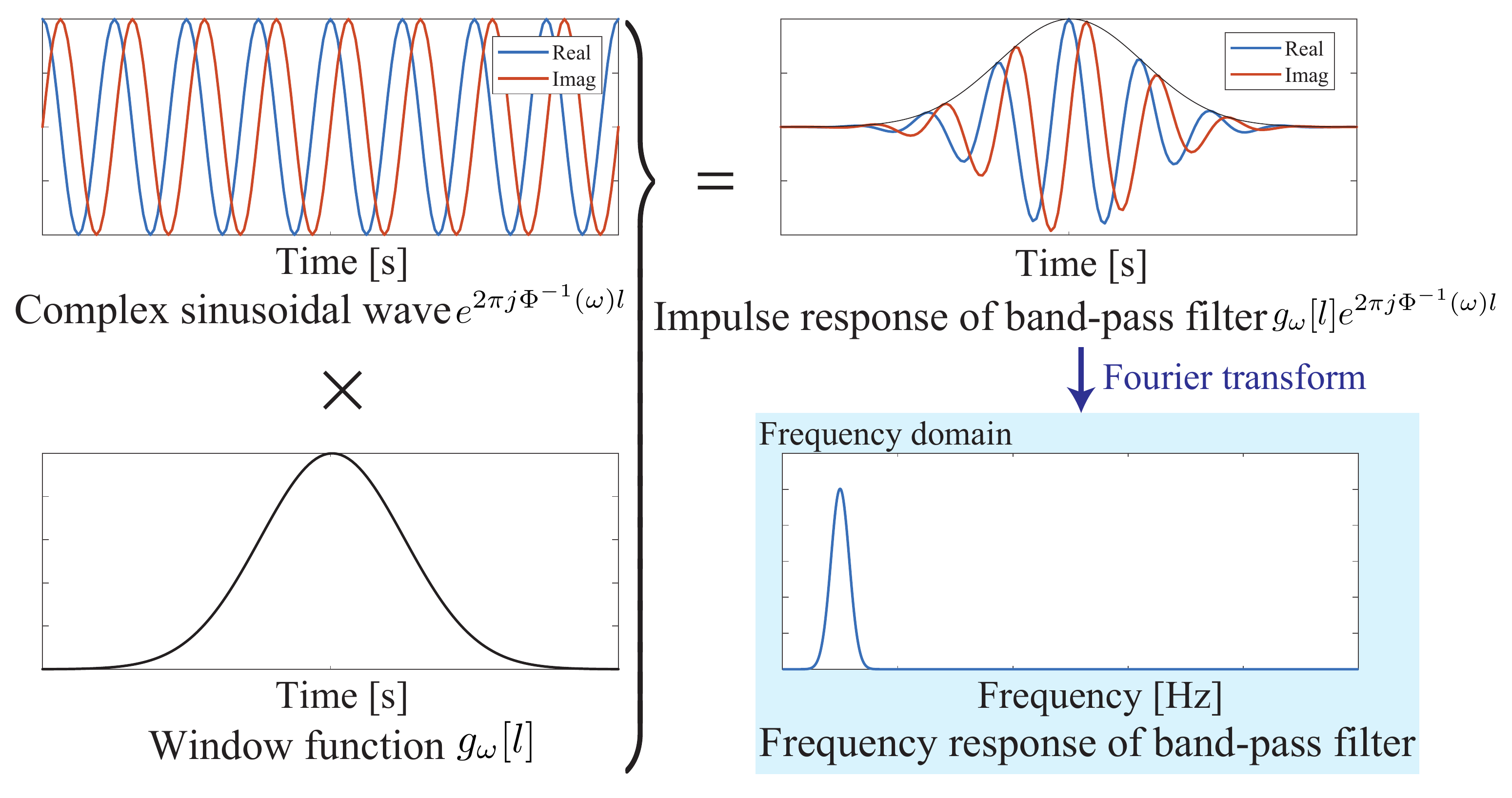}
\vspace{-5pt}
\caption{Illustration of band-pass filter $g_\omega[l]\,e^{2 \pi j \Phi^{-1}(\omega) l}$ in Eq.~\eqref{eq:wfbf}.}
\label{fig:convF}
\vspace{-3pt}
\end{figure}

The WFBF is a PRFB whose frequency scale can be defined by a user.
The WFBF can be written as the following form:
\vspace{-2pt}
\begin{equation}
\vspace{-2pt}
\mathscr{F}(\bm{x})[\omega,k]
= \sum\limits_{l = 0}^{L-1} x[l-a_{\omega}k]\,
\overline{g_{\omega}[l] \, e^{2 \pi j \Phi^{-1}(\omega) l}},
\label{eq:wfbf}
\end{equation}
where $\Phi$ is a frequency-warping function which is defined so that the resulting WFBF has the desired frequency scale (see \cite{holighaus2014class,holighaus2015designing} for the regularity required for $\Phi$ such as $\mathcal{C}^1$-diffeomorphism and having positive derivative), and the parameters with the subscript $\omega$ may be different for each frequency band.
Since Eq.~\eqref{eq:wfbf} is the downsampled convolution between the signal and $g_\omega[l]\,e^{2 \pi j \Phi^{-1}(\omega) l}$ (see Fig.~\ref{fig:convF}) which can be computed efficiently via the fast Fourier transform (FFT), WFBF is a collection of bandpass filters whose center frequencies are decided by the warping function, and the window functions are automatically derived according to the design requirement \cite{holighaus2014class,holighaus2015designing}.
As some illustrative examples,
\begin{equation}
    \Phi_{\rm STFT}(\omega) = \omega/b, \qquad \Phi_{\rm wavelet}(\omega) = \log_c(\omega),
\end{equation}
recover STFT and wavelet transform, respectively.
By only defining the warping function, one can easily realize WFBF with the desired frequency scale through implementation in the LTFAT toolbox \cite{ltfatnote030}.

\begin{figure}[t]
\centering 
\includegraphics[width = 0.97\columnwidth]{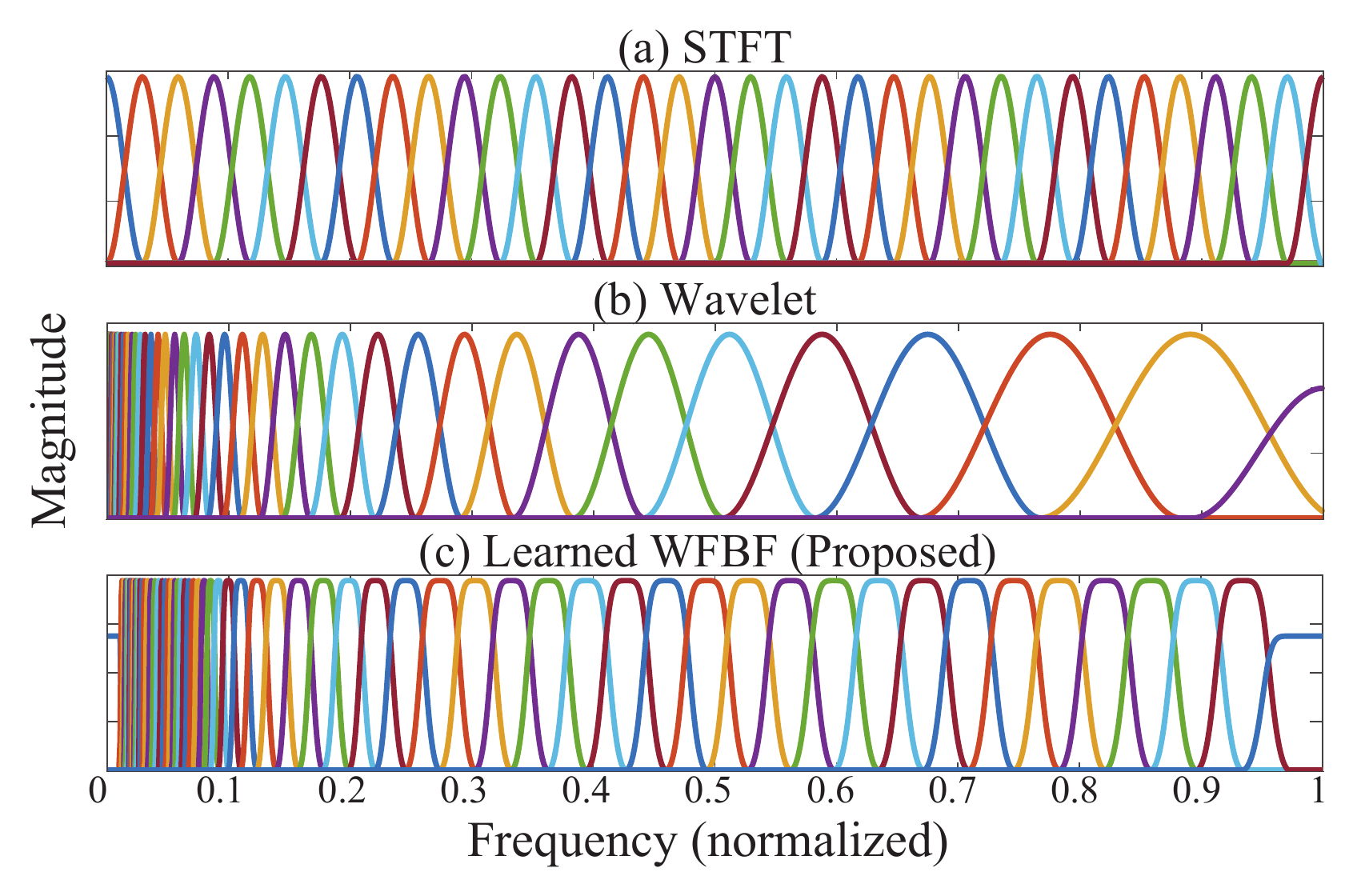}
\vspace{-5pt}
\caption{Frequency responses of (a) STFT, (b) wavelet transform, and (c) the proposed WFBF learned from training dataset ($f_{s}\!=\!16$ kHz).}
\label{fig:est_init}
\vspace{-5pt}
\end{figure}

\label{sec:exp}

\mysubsection{Proposed sound source enhancement in WFBF domain}

T-F masking is applied in WFBF domain instead of STFT domain as
\begin{equation}
    \hat{\mathscr{F}(\bm{s})}[\omega,k] = 
    \mathcal{M}_\theta^{\mathscr{F}}\!(\Psi_{\!\mathscr{F}})[\omega,k] \, 
    \mathscr{F}(\bm{x})[\omega,k],
\end{equation}
where $\hat{\mathscr{F}(\bm{s})}[\omega,k]$ is the estimated target signal, $\mathcal{M}_{\theta}^{\mathscr{F}\!}$ is a DNN-based regression function, and $\Psi_{\!\mathscr{F}}$ is the input feature.
After masking, the estimated time-domain signal is recovered from $\hat{\mathscr{F}(\bm{s})}[\omega,k]$ thanks to the perfect reconstruction property of WFBF.

In order to design the warping function, we propose to use the frequency-wise energy of a training dataset.
Since MSE assumes that the error of all frequency has uniform variance, T-F transform should be designed to fulfill this assumption.
To do so, the error of T-F masking by the oracle PSM in Eq.~\eqref{eq:errorOraPSM} is collected from the training dataset of $\mathcal{M}_{\theta}^{\mathscr{F}\!}$ because that is what the training tries to minimize.
Then, its power spectral density (PSD) is estimated from the corrected error, which should be normalized to met the assumption of MSE.
For the normalization, we propose to obtain the warping function $\Phi$ by
\begin{equation}
    \Phi_{\rm Prop} = {\rm cumsum}(\bm{\sigma} + \lambda),
    \label{eq:propPhi}
\end{equation}
where $\bm{\sigma}$ is a vector of the error PSD, $\lambda\geq0$ is a small regularization parameter, and ${\rm cumsum}$ is the cumulative sum taken from lower to higher frequency.
The role of $\lambda$ is to avoid an excessively wide frequency band, where a larger $\lambda$ makes WFBF closer to STFT.

Using the proposed PRFB based on WFBF, MSE can be appropriately used as the cost function without the normalization weight,
\begin{equation}
     \mathcal{J}_{\rm Prop}(\theta) \!=\!\! 
     \sum\limits_{\omega = 1}^{\Omega_{\!\mathscr{F}}}
     \!\sum\limits_{k = 1}^{K_{\!\mathscr{F}}}
     \!\left|
     \mathcal{M}_{\theta}^{\mathscr{F}}\!(\Psi_{\!\mathscr{F}})[\omega,k]
      \,\mathscr{F}\!(\bm{x})[\omega,k] 
      -
      {\mathscr{F}\!(\bm{s})}[\omega,k]
     \right|^2.
     \label{eq:costProposed}
\end{equation}
The assumption mismatch is removed by training a DNN based on this cost function because the energy is normalized for each frequency by the filterbank, and its gradient is stably balanced as the weight is removed.
The proposed PRFB can be viewed as the pre-emphasis which emphasize the important frequency range and de-emphasize unimportant range that are learned from the training data.

\begin{table}[t]
\centering
\caption{Network architectures.}
\vspace{2pt}
\begin{tabular}{l | c | l} \hline \hline
Layer num. & Type & Size (activation) \\ \hline
Input & Fully & 64/128/257 $\to$ 512 (ReLU)\\ 
Hidden 1 & BLSTM & 512 $\to$ 512\\
Hidden 2 & BLSTM & 512 $\to$ 512\\
Output & Fully & 512 $\to$ 64/128/257 (sigmoid)\\ 
\hline \hline
\end{tabular}
\label{tab:DNNachi}
\vspace{-8pt}
\end{table}

{
\renewcommand\arraystretch{1.08}
\begin{table*}[!t]
\centering 
\caption{
SDR of enhanced speech.
The bold fonts indicate the highest score, while asterisks indicate significance.
}
\vspace{2pt}
\scriptsize
\begin{tabular}{ @{}C{0.25}@{} | @{}C{0.27}@{} | @{}C{0.25}@{} | @{}C{0.29}@{} | 
                 @{\hspace{5pt}}C{0.23}@{} @{}C{0.23}@{} @{}C{0.23}@{} @{}C{0.23}@{\hspace{5pt}} } \hline \hline
Input SNR & Input dimension & T-F transform & Cost function 
& {\it factory1} &{\it factory2} & {\it f16} &{\it babble} \\
\hline \hline
\multirow{8}*{6~dB}&    & STFT & MSE & 10.48 & 12.60 & 10.50 & 9.71  \\  
                   &64  & STFT & weighted MSE & 9.87 & 11.86 & 10.31 & 8.90  \\ \cline{3-8}
                   &    & Proposed & MSE & $^*${\bf 12.21}& $^*${\bf 14.65}& $^*${\bf 12.58}& $^*${\bf 11.66} \\ \cline{2-8}
                   
                   &    & STFT & MSE & 10.67 & 13.16 & 10.82 & 9.92  \\  
                   &128 & STFT & weighted MSE & 9.77 & 11.92 & 10.46 & 8.89  \\ \cline{3-8}  
                   &    & Proposed & MSE & $^*${\bf 11.82}& $^*${\bf 14.17}& $^*${\bf 12.24}& $^*${\bf 11.13}  \\ \cline{2-8}
                   
                   &\multirow{2}*{257} & STFT & MSE & 10.07 & 12.44 & 10.33 & 9.53\\ 
                   &     & STFT & weighted MSE & 9.50 & 11.44 & 10.22 & 8.81 \\                   
                   
\hline \hline

\multirow{8}*{0~dB}&    & STFT & MSE & 6.35 & 8.63 & 6.39 & 5.41  \\  
                  &64  & STFT & weighted MSE & 5.65 & 7.51 & 6.19 & 4.72  \\ \cline{3-8}
                  &    & Proposed & MSE & $^*${\bf 8.11}& $^*${\bf 10.71}& $^*${\bf 8.62}& $^*${\bf 7.27} \\ \cline{2-8}
                  
                  &    & STFT & MSE &  6.31 & 8.87 & 6.49 & 5.45 \\  
                  &128 & STFT & weighted MSE & 5.37 & 7.31 & 6.27 & 4.49  \\ \cline{3-8}  
                  &    & Proposed & MSE & $^*${\bf 7.85}& $^*${\bf 10.40}& $^*${\bf 8.42}& $^*${\bf 6.92}  \\ \cline{2-8}
                   
                   &\multirow{2}*{257} & STFT & MSE & 5.58 & 8.24 & 6.02 & 5.03\\ 
                   &     & STFT & weighted MSE & 5.04 & 6.99 & 5.93 & 4.43 \\  
\hline \hline

\multirow{8}*{-6~dB}&    & STFT & MSE & 3.04 & 5.03 & 3.17 & 1.98\\  
                  &64  & STFT & weighted MSE & 2.66 & 4.04 & 3.02 & 1.75  \\ \cline{3-8}
                  &    & Proposed & MSE & $^*${\bf 4.17}& $^*${\bf 6.74}& $^*${\bf 4.76}& $^*${\bf 2.99} \\ \cline{2-8}
                  
                  &    & STFT & MSE & 2.90 & 5.10 & 3.20 & 1.96  \\  
                  &128 & STFT & weighted MSE & 2.33 & 3.78 & 3.15 & 1.55 \\ \cline{3-8}  
                  &    & Proposed & MSE & $^*${\bf 4.09}& $^*${\bf 6.65}& $^*${\bf 4.78}& $^*${\bf 2.83}  \\ \cline{2-8}
                   
                   &\multirow{2}*{257} & STFT & MSE & 2.30 & 4.64 & 2.87 & 1.77\\ 
                   &     & STFT & weighted MSE & 2.02 & 3.54 & 2.80 & 1.58 \\  
\hline \hline
\end{tabular}
\label{tab:6dB}
\vspace{-2pt}
\end{table*}
}

\mysection{Experiment}
In order to confirm the correctness of the proposed method, the performance of SSE is investigated by comparing the STFT-domain PSM in Eq.~\eqref{eq:usualMSE} \cite{Erdogan2015}, that with the weighted MSE in Eq.~\eqref{eq:weightedLS}, and the proposed method in Eq.~\eqref{eq:costProposed}.

Before that, as a preliminary experiment, frequency-wise variance of the error $\varepsilon$ in Eq.~\eqref{eq:errorOraPSM} was calculated to see how the proposed warping function $\Phi_{\rm Prop}$ in Eq.~\eqref{eq:propPhi} works.
By using the training dataset explained below, the error of the oracle PSM in Eq.~\eqref{eq:errorOraPSM} was collected, and its PSD was estimated by the Welch estimator.
The frequency response of the filterbank obtained by the proposed method $(\lambda\!=\!0.1)$ is shown in Fig.~\ref{fig:est_init}, where an example of the obtained WFBF-domain representation can be found at the bottom of Fig.~\ref{fig:propIm}.
The learned WFBF has a characteristic in between STFT and wavelet transform, which analyzes around 100--800 Hz finely and other bands coarsely.
The variances of the error $\varepsilon$ for each frequency are shown in Fig.~\ref{fig:preEx}.
From the figure, it can be seen that the proposed filterbank obtained more balanced distribution of the masking error.
While smaller $\lambda$ results in more balanced variance distribution, wider frequency band appears in high-frequency range which is not favorable.
We chose $\lambda = 0.1$ to balance this trade-off.
Note that computation of the proposed design method is quite fast and scalable because it only requires PSD estimation of masking error which can be computed by STFT with FFT.

\mysubsection{Experimental conditions}
\vspace{5pt}
\mysubsubsection{Dataset}
The Wall Street Journal (WSJ-0) corpus and noise dataset CHiME-3 were used as the training datasets of the target source and noise, respectively. 
The WSJ-0 dataset consisted of 14\,633 utterances. 
The utterances were randomly separated into two sets: a training set consisting of 13\,170 speech files and validation set including 1463 speech files. CHiME-3 consisted of four types of background noise: \textit{cafes}, \textit{street junctions}, \textit{public transport (buses)}, and \textit{pedestrian areas} \cite{Barker2015}. 
The noisy signals for the training/validation dataset were formed by mixing clean speech utterances with the noise at signal-to-noise ratio (SNR) levels of -6 to 12 dB. 
As the test datasets, 500 utterances randomly selected from the TIMIT corpus were used for the target-source dataset, and four types of ambient noise {\it factory1}, {\it factory 2}, {\it f16}, and {\it babble} from the NOISEX92 dataset were used as the noise dataset. 
All files were recorded at sampling rate of 16 kHz. 

\vspace{-5pt}
\mysubsubsection{DNN architecture and setup}
The performances of SSE in the proposed WFBF domain and the conventional STFT domain were compared on the network with the two bidirectional long short-term memory (BLSTM) consisting of 512 cells.
As the activation function of the input layer, the rectified linear unit (ReLU) was used.
The sigmoid function was used at the output layer for limiting the values within the range 0 to 1.
In the proposed method, input acoustic feature $\Psi_{\!\mathscr{F}}$ was defined as
\begin{equation}
    \Psi_{\!\mathscr{F}}[\omega,k] = \ln(|\mathscr{F}(\bm{x})[\omega,k]|),
\end{equation}
where $|\cdot|$ denotes the absolute value.
In the proposed method, the number of frequency bins (and thus input dimension) was set to 64/128.
In the conventional STFT, the number of frequency bins was set to 512 (i.e., input dimension was 257), and the window is shifted by 256 samples.
To match the input dimensions between the conventional and proposed methods, 64/128 dimensional log-mel transform matrix and its pseudo-inverse were applied to STFT, i.e., input feature of the conventional STFT $\Psi$ was
\begin{equation}
    \Psi[\omega,k] = \ln({\rm Mel}[|X[\omega,k]|]),
\end{equation}
where ${\rm Mel}[\cdot]$ denotes the 
mel matrix multiplication.
The conventional method without log-mel transform is also considered as a baseline.
In summary, the input dimensions of the conventional method were 64, 128, and 257.
These architectures are summarized in Table~\ref{tab:DNNachi}.
They were trained 200 epochs in the same way as \cite{Erdogan2018}, where each epoch contained 1000 utterances, and mini-batch size was 5.
The learning and dropout rates were decreased linearly.

\vspace{-1pt}
\mysubsection{Experimental result}
The performances of SSE were measured by the signal-to-distortion ratio (SDR).
The experimental results are summarized in Table~\ref{tab:6dB}.
Bold font indicates the best score within the same condition, and asterisks represent that the scores were significantly higher than those of second-placed methods within the same condition (provided by the paired one-sided $t$-test with $p\!<\!0.01$).
For all cases, the proposed WFBF achieved the highest scores.
The scores of the conventional STFT tends to decrease as the input dimension decreases, while those of the proposed WFBF did not.
It was also confirmed that weighted MSE in STFT domain obtained less performances than the usual MSE, which should be because of the difficulty of the optimization.
These results indicate that using the learned T-F transform instead of the ordinary STFT is more efficient for SSE.
Note again that, after calculating the proposed warping function in
Eq.~\eqref{eq:propPhi}, the WFBF can be easily obtained through
\texttt{warpedfilters} function in the LTFAT toolbox \cite{ltfatnote030} by
just substituting it.

\vspace{-1pt}
\mysection{Conclusions} 
\vspace{-1pt}
In this paper, a data-driven design method of PRFB using WFBF was proposed for DNN-based SSE.
By considering WFBF, the learning problem of T-F representation was reduced to calculation of the one-dimensional frequency-warping function, which is obtained through PSD of oracle masking error.
Since the calculation of the proposed warping is cheap, it can be easily adapted to different dataset in contrast to a fully DNN-based learning of T-F-like representation.
Future works include simultaneous optimization of a T-F mask estimator and the warping function as well as consideration of a better DNN architecture suitable for processing in the filterbank domain.

\pagebreak
\label{sec:refs}
\clearpage
\ninept

\end{document}